\documentstyle[aps,prl,epsf,twocolumn,floats]{revtex}
\begin{document}
\twocolumn[\hsize\textwidth\columnwidth\hsize\csname@twocolumnfalse\endcsname
\title{
Locally critical point in an anisotropic Kondo lattice}
\author{D. R. Grempel$^{(a)}$ and Qimiao Si$^{(b)}$}
\address{$^{(a)}$CEA-Saclay/DRECAM/SPCSI, 91191 Gif-sur-Yvette, France
\\
$^{(b)}$Department of Physics \& Astronomy, Rice University, Houston,
TX 77005--1892, USA}

\maketitle

\begin{abstract}
We report the first numerical identification of a locally quantum 
critical point, at which the criticality of the local Kondo physics
is embedded in that associated with 
a magnetic ordering. 
We are able to numerically access the quantum critical behavior
by focusing on a Kondo-lattice model with Ising 
anisotropy.
We also establish  that the critical exponent
for the ${\bf q}-$dependent dynamical spin susceptibility
is fractional and compares well with 
the experimental value for heavy 
fermions.
\end{abstract}
]

\narrowtext

How to properly describe heavy fermion metals near
quantum critical points (QCPs) is a subject
of intensive current research.
It is well established experimentally~\cite{StewartRMP} 
that these systems 
are
prototypes of non-Fermi liquid metals\cite{VarmaNFL,questions}.
In a number of cases, striking deviations from
the commonly applied $T=0$ spin-density-wave picture
(usually referred to as the Hertz-Millis picture)
~\cite{Sachdev-book} have been seen.
In particular, experiments\cite{Schroder2,Stockert,Gegenwart,Ishida,Montfrooij}
have shown that the spin dynamics in 
the quantum critical 
regime can display fractional exponents essentially everywhere 
in the Brillouin zone, as well as $\omega/T$ scaling. These 
features are completely unexpected in the Hertz-Millis picture -- which
corresponds to a Gaussian fixed point -- and they 
directly imply the existence of quantum critical metals that have
to be described by an interacting fixed point.
A number of theoretical approaches are being undertaken to search
for such non-Gaussian quantum critical
metals\cite{lcp-nature,lcp-long,Coleman,Sachdev}.

Here we are concerned with a new class of QCP
\onlinecite{lcp-nature,lcp-long},
which has properties that bear a close
similarity to those seen experimentally. The key difference between
the traditional Hertz-Millis QCP and this locally critical point (LCP)
is that, in the latter,  the local Kondo physics  itself becomes critical
at the antiferromagnetic ordering transition. 
Such a LCP was shown to arise
in an extended dynamical mean field theory 
(EDMFT) of a Kondo lattice model.
The latter was mapped onto a self-consistent impurity model 
-- the Bose-Fermi Kondo model -- which in turn was analyzed using
a renormalization-group (RG) approach, based on an
$\epsilon$ expansion.

One of the key issues is whether 
the destruction of the Kondo effect is accompanied by
a fractional exponent in the frequency/temperature dependences of the 
dynamical spin susceptibility.
The fractional nature of the exponent has been seen experimentally,
and is characteristic of a non-Gaussian magnetic 
quantum critical metal. The calculation of this exponent requires
a detailed numerical
study of the quantum critical behavior.

In this paper we demonstrate that, 
for an anisotropic version of the Kondo lattice model,
the EDMFT equations can be solved using a Quantum Monte
Carlo (QMC) method~\cite{Grempel-Rozenberg}, including 
at its QCP. 
We also analyze them using a
saddle-point approximation
\cite{Sengupta-Georges,SachdevRead}.

We consider the following Kondo lattice model:
\begin{eqnarray}
{\cal H}
= \sum_{ ij\sigma} t_{ij} ~c_{i\sigma}^{\dagger} c_{j\sigma}
+ \sum_i J_K ~{\bf S}_{i} \cdot {\bf s}_{c,i}
+ \sum_{ ij} (I_{ij} /2)
~S_{i}^z S_{j}^z .
\label{kondo-lattice}
\end{eqnarray}
Here, $t_{ij}$ specify the bandstructure ($\epsilon_{\bf k}$)
and density of states [$\rho_0(\epsilon)$]
of the conduction $c-$ electrons,
$J_K$ describes the on-site Kondo coupling between
a spin-${1 \over 2}$ local moment ${\bf S}_{i}$ and a
conduction-electron spin density ${\bf s}_{c,i}$ and,
finally, $I_{ij}$ denote the exchange interactions between
the $z-$components of two local moments.

We study this model using the EDMFT 
approach~\cite{Smith1,Smith2,Chitra}, which maps the
model onto a self-consistent anisotropic Bose-Fermi Kondo model:
\begin{eqnarray}
{\cal H}_{\text{imp}}
=&& J_K ~{\bf S} \cdot {\bf s}_c
+ \sum_{p,\sigma} E_{p}~c_{p\sigma}^{\dagger}~ c_{p\sigma}
\nonumber\\ &&
+ \; g \sum_{p} S^z 
\left( \phi_{p} + \phi_{-p}^{\;\dagger} \right)
+ \sum_{p} w_{p}\,\phi_{p}^{\;\dagger} {\phi}_{p}\;,
\label{H-imp}
\end{eqnarray}
where the parameters $E_p$, $w_p$ and $g$ are determined by a set of 
 self-consistency equations.
The latter are dictated by the translational
invariance:
\begin{eqnarray}
G_{\text{loc}} (\omega) &=& \sum_{\bf k} G( {\bf k}, \omega )
= \int_{-t}^t d \epsilon {\rho_0(\epsilon)
\over {\omega + \mu -\epsilon - \Sigma(\omega)}} ,
\nonumber \\[-1ex]
\label{self-consistent} \\[-1ex]
\chi_{\text{loc}} (\omega) &=& \sum_{\bf q} \chi ( {\bf q}, \omega ) 
= \int_{-I}^I d \epsilon {\rho_I(\epsilon)
\over {M(\omega)+ \epsilon}},
\nonumber
\end{eqnarray}
where $G_{\text{loc}} (\omega)$ is the local conduction-electron Green
function and $\chi_{\text{loc}}
(\tau)= <T_{\tau} S^z(\tau) S^z(0)>$ is
 the local spin susceptibility. The
conduction-electron
 and spin self-energies are 
$\Sigma(\omega)$ and $M(\omega)$,
respectively, with  
\begin{eqnarray}
M(\omega) = \chi_0^{-1}(\omega) + {1 \over \chi_{
\text{loc}}(\omega)}\;,
\label{M}
\end{eqnarray}
where the Weiss field $\chi_0^{-1}$ characterizes the 
boson bath,
\begin{eqnarray}
\chi_{0}^{-1}(i\omega_n)
= - g^{2} \sum_{p}  
2 w_{p} / [(i\omega_n)^{2} - w_{p}^{2}]\;.
\label{chi_0}
\end{eqnarray} 
It has been recognized\cite{lcp-nature,lcp-long}
that the solution depends crucially 
on the dimensionality of the spin fluctuations,
which 
enters through
the RKKY-density-of-states:
\begin{eqnarray}
\rho_{I} (\epsilon) \equiv  \sum_{\bf q} \delta
( \epsilon  - I_{\bf q} ) .
\label{rkky-dos}
\end{eqnarray}
In this paper, we will consider 
the following specific form, characteristic of two-dimensional
fluctuations:
\begin{eqnarray}
\rho_{I} (\epsilon) = 
(1/{2 I}) \Theta(I - | \epsilon | ) ,
\label{rkky-dos-2D}
\end{eqnarray}
where $\Theta$ is the Heaviside function.
Enforcement of the self-consistency condition on the
conduction-electron Green function is
not essential for the critical
properties discussed here,
provided that the density of states at the 
chemical potential ($\mu$)
is finite;
the corresponding bath density of states
$\sum_{p} \delta (\omega - E_{p}) = N_0$
is also finite.

If, instead of 
the general 
form 
Eq.~(\ref{rkky-dos})
we were to choose a semi-circular RKKY-density-of-states
(representative of the 3D case), our EDMFT  equations would 
become essentially the same as those of 
Refs.~\onlinecite{Grempel-Rozenberg,Sengupta-Georges,SachdevRead,Kajueter}.

To cast the impurity model
in the form
of a functional integral,
we adopt a
bosonized 
representation of the fermionic bath,
$c_{\sigma}^{\dagger} = F_{\sigma}^{\dagger} { 1 \over \sqrt{2\pi a}}
{\rm e}^{i\Phi_{\sigma}}\;,$
and
a canonical transformation using
$U={\rm exp}[-i(\sqrt{2}/2)\Phi_s(\Sigma_{\sigma}
\sigma X_{\sigma\sigma})]$,
where~\cite{Kotliar-Si}
$X_{\alpha\beta} \equiv |\alpha><\beta|$ are the Hubbard 
operators and 
$\Phi_s \equiv (\Phi_{\uparrow}-\Phi_{\downarrow})/\sqrt{2}$,
leading to 
\begin{eqnarray}
{\cal H}_{\text{imp}}'
\equiv&& U^{\dagger}{\cal H}_{\rm imp} U
=
{\cal H}_0(\phi_s,\phi_c) + 
\Gamma
S^x 
+ \Gamma_z S^z s^z
\nonumber\\ &&
+ \;  g \sum_{p} S^z 
\left( \phi_{p} + \phi_{-p}^{\;\dagger} \right)
+ \sum_{p} w_{p}\,\phi_{p}^{\;\dagger} {\phi}_{p}\;.
\label{H-imp'}
\end{eqnarray}
Here, 
${\cal H}_0(\phi_s,\phi_c)$ describes the bosonized
conduction electron bath,
$S^x \equiv ( X_{\uparrow \downarrow} 
F_{\downarrow}^{\dagger} F_{\uparrow} + H.c. )/2 $,
$s^z = \left ({{d\Phi_s}\over {d x}} \right )_{x=0} 
{1 \over {2\pi}} $ is the 
spin density of the conduction electron
bath,
$\Gamma = {{\sqrt{2}J_{K}} \over{4\pi a}}$,
and ${\Gamma_z} = {{2 \sqrt{2}}
\over {\pi N_0} } \left [ \tan^{-1}
\left( {{\pi N_0 J_K} \over {4}} \right) 
- {\pi \over 2} \right ]$.
($\Gamma$ and ${\Gamma_z}$ can vary independently 
when 
we allow the longitudinal and spin-flip parts of the Kondo
interaction to be different.)
Eq.~(\ref{H-imp'})
describes an Ising spin in a transverse field 
$\Gamma $, with a retarded 
self-interaction 
that is long-ranged in time~\cite{Grempel-Rozenberg}.
The associated partition function 
is 
\begin{eqnarray}
Z_{\text{imp}}'&&
\sim \int {\cal D} n \;d \lambda 
{\rm exp} [ 
-{1 \over 2} \int_0^{\beta} d \tau [ 
{i\lambda } (n^2 -{1 \over 4} ) 
+ { 1 \over {g_c}} 
(\partial_{\tau} n)^2 
\nonumber\\ 
&&-\int_0^{\beta} d \tau' n(\tau) n(\tau')
( \chi_0^{-1}(\tau-\tau') 
- {\cal K}_c(\tau-\tau') )
]
]
\;.
\label{Z-imp}
\end{eqnarray}
Here, $g_c \sim (1/N_0)[\ln{2 \over {N_0J_{K}}} ]^{-1}$
and ${\cal K}_c$ comes
from integrating out the
electron bath
($\kappa_c \sim \Gamma_z^2N_0^2$),
\begin{eqnarray}
{\cal K}_c(i\omega_n) = \kappa_c |\omega_n| .
\label{K}
\end{eqnarray} 

To first 
gain qualitative insights,
we carry out a saddle-point analysis of Eq.~(\ref{Z-imp}).
This analysis 
 is formally exact
when the number of components for the field $n$ is generalized from
1 to $N$ and a large-$N$ limit is subsequently 
taken~\cite{Sengupta-Georges,SachdevRead}.
At the saddle point level, $i\lambda = \lambda_0$,
and 
\begin{eqnarray}
\chi_{\text{loc}} (i\omega_n)
= \left [ \lambda_0 - \chi_0^{-1}(i\omega_n) + \kappa_c |\omega_n|
+ \omega_n^2/g_c \right ]^{-1} .
\label{chi-loc}
\end{eqnarray} 
There is also a constraint equation that reads
\begin{eqnarray}
(1/ \beta) \sum_{\omega_n} 
\chi_{\text{loc}} (i \omega_n) {\rm e}^{i\omega_n 0^+} = 1/4\;.
\label{constraint}
\end{eqnarray} 
The QCP is reached when the 
static susceptibility at the 
ordering wavevector ${\bf Q}$, specified by $I_{\bf Q} = -I$,
becomes divergent. Given that\cite{Smith1,Smith2,Chitra}
\begin{eqnarray}
\chi ({\bf q},\omega) = [M(\omega) + I_{\bf q}]^{-1},
\label{chi-Q}
\end{eqnarray} 
it 
implies $M(\omega=0)\to I$.
This, in turn, establishes that $\chi_{\text{loc}}(\omega =0)$
is also divergent [through 
Eqs.~(\ref{self-consistent},\ref{rkky-dos-2D})]
and that $\lambda_0 = \chi_0^{-1}(0)= I$
[from Eqs.~(\ref{chi-loc},\ref{M})].
The self-consistency equation~(\ref{self-consistent}) then
simplifies and becomes
\begin{eqnarray}
\chi_0^{-1} (\omega) + { 1 \over {\chi_{\text{loc}}(\omega)}}
&& = M(\omega) \nonumber \\
&&= I + 2I \exp
\left[ -2I  {\chi_{\text{loc}}(\omega)} \right] .
\label{self-consistent2D-2}
\end{eqnarray}
We 
immediately 
see, from 
Eqs.~(\ref{chi-loc},\ref{constraint},\ref{self-consistent2D-2}), 
that
\begin{eqnarray}
\chi_0^{-1} (\omega) &&
= I - 2  \Lambda \ln^{-1}\left({\Lambda
\over -i\omega}\right)\;, 
\label{solution-leading-1}\\
\chi_{\text{loc}} (\omega) 
&& = {1 \over { 2 \Lambda}}
\ln\left({\Lambda \over {-i\omega} }\right)\;,
\label{solution-leading-2}
\end{eqnarray} 
are self-consistent to the leading order.

Since the $\omega$-dependent part of $M(\omega)$ 
is subleading compared to both $1/\chi_{\text{loc}}(\omega)$ and 
the $\omega$-dependent part of $\chi_0^{-1}(\omega)$, 
we can determine the spin self-energy entirely in terms of the leading
order solution for $\chi_{\text{loc}}
(\omega)$ in Eq.~(\ref{solution-leading-2})
and the second equality in Eq.~(\ref{self-consistent2D-2}) :
\begin{eqnarray}
M (\omega) = I + 2I (-i\omega/\Lambda)^{\alpha}\;,
\label{solution-leading-4}
\end{eqnarray}
with the exponent $\alpha = I/\Lambda$.
Now, to satisfy the first equality of 
Eq. (\ref{self-consistent2D-2}), we need to add 
sub-leading terms to the Weiss field and replace
 Eq. (\ref{solution-leading-1}) by
\begin{eqnarray}
\chi_0^{-1} ( \omega) = I -  2 \Lambda \ln^{-1}\left({\Lambda
\over {-i\omega}}\right) - c (-i\omega)^{\alpha}\;.
\label{chi-0-subleading}
\end{eqnarray}
Eq.~(\ref{chi-loc}) then leads to:
\begin{eqnarray}
{1 \over {\chi_{\text{loc}} (\omega)}}
= 2 \Lambda \ln^{-1}\left({\Lambda \over {-i\omega}}\right)
+ c' (-i\omega)^\alpha -  i \kappa_c \omega 
\label{chi-loc-subleading}
\end{eqnarray}
At the saddle point level, however, 
the critical amplitudes for the local susceptibility
are pinned to the initial parameters
of the Weiss
field\cite{amplitude-pinning}.
The self-consistent
amplitudes of $\chi_{\rm loc}$ is not determined from the 
leading terms alone and $c'=c$;
a fractional $\alpha$ cannot
emerge.
We now show that
a fractional $\alpha$ does arise\cite{amplitude-pinning}
in the physical case
corresponding to Eq.~(\ref{H-imp'}).

\begin{figure}[t]
\centerline{\epsfxsize=75mm\epsfbox{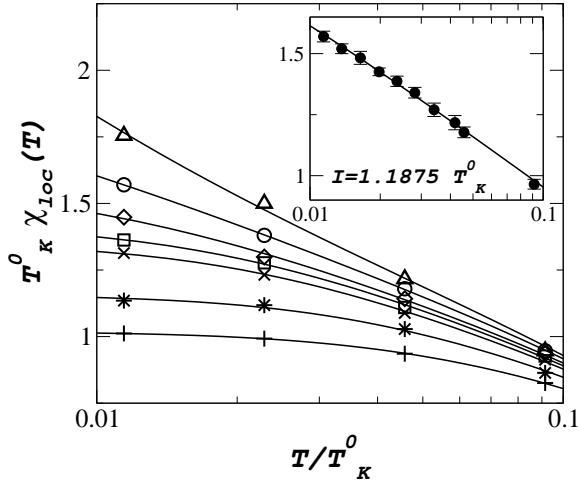}}
\vspace{1ex}
\caption{%
Static local susceptibility vs. $T$
for $I/T^0_K$ = 0 (+), 0.75 ($\ast$), 1 ($\times$),
1.0625 ($\Box$), 1.125 ($\Diamond$), 1.1875 (O) and 1.25
($\triangle$). The solid lines are the fits described in the
text. Inset: results for $I = 1.1875\;T^0_K$,
with $\delta(T=0) = 5\times10^{-3}$.
}
\label{fig1}
\end{figure}
We have directly studied the 
physically relevant 
case [Eq.~(\ref{H-imp'})] 
numerically
using the QMC algorithm of
Refs.~\onlinecite{Grempel-Rozenberg,qmc}.
Starting from a
trial 
$\chi_0^{-1}(\tau)$, we compute 
$\chi_{\text{loc}}(\tau)$. A new $\chi_0^{-1}(\tau)$ 
is then obtained from
the self-consistency
Eq.~(\ref{self-consistent}),
and the process is repeated until
convergence is achieved.
The results reported below are obtained for $\kappa_c = \pi$
[cf. Eq.~(\ref{K})]. In this special case, the model can
be solved exactly at $I=0$, providing a check on the QMC
algorithm~\cite{Grempel-Rozenberg}. 
We choose 
$N_0 \Gamma = 0.19$ 
such that the 
bare
Kondo scale
(the inverse of
the static local susceptibility at $I=0$), $ N_0 T^0_K = 0.17$,
 is well above the lowest
temperature attainable in our simulations, 
$T_{min} = 10^{-2}\;T^0_K$. The number of Trotter time slices used varies
between 64 (for the higher temperatures) and 512 (for the lowest one).
We perform between $10^4$ and $10^6$ MC steps per time slice and
between five and twenty self-consistency iterations. 
The numerical error is controlled by the last step.
We estimate that the calculated susceptibilities
are accurate to within 3\%.  

Fig.~\ref{fig1} shows the temperature dependence of 
$\chi_{\text{loc}}(\omega= 0)$ for several values
of $I$.  The results exhibit  
very little $I$-dependence above $T \approx 0.1\;T^0_K$. 
At lower temperatures,
a saturation of $\chi_{\text{loc}}$ 
at an $I$-dependent value is seen in the five
lower curves. The two upper curves
show no saturation;
instead, the results are consistent with a logarithmic
$T$-dependence as shown in the inset of Fig.~\ref{fig1},
which contains data at many more points of temperature.
\begin{figure}[t]
\centerline{\epsfxsize=90mm\epsfbox{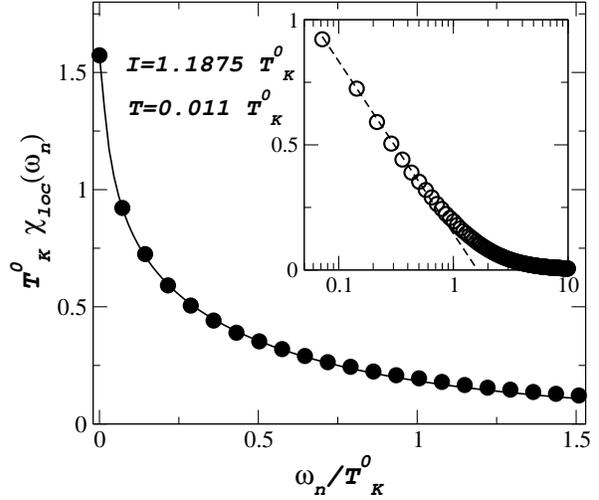}}
\vspace{1ex}
\caption{%
Matsubara-frequency dependence of $\chi_{\text{loc}}$ 
at $T=0.011\;T^0_K$ and 
$I = 1.1875 \;T^0_K$.
The solid line is the fit obtained from
the Hilbert transform of 
Eq.~(\ref{fitting-function}). Inset: the same data on a logarithmic 
 plot with the point at $\omega_n =0$ excluded. The dashed line is
the predicted critical behavior~[10,11] 
with $\alpha = 0.72$ and $\Lambda =1.54\;T^0_K$.}
\label{fig2}
\end{figure}
To interpret these numerical results,
and inspired by the arguments developed earlier,
we use the following fitting function for
$\chi_{\text{loc}}''(\omega)$:
\begin{equation}
\chi_{\text{loc}}''(\omega) = {\alpha \over 2 I} 
\Theta\left(\Lambda - |\omega|\right)\;
{\pi \over 2}
\tanh\left({\omega \over \Lambda\;\delta}\right)\;,
\label{fitting-function}
\end{equation} 
where
$\alpha$,  
$\Lambda$,
and $\delta$ are fitting
 parameters that may depend on $I$ and $T$.
The parameter
 $\delta$ is a measure of the proximity to the QCP as  
 Eq.~(\ref{fitting-function}) implies that 
 $\chi^{-1}({\bf Q}) = 2 I \delta^{\alpha}$ for $\delta \ll 1$.
By Hilbert transforming Eq.~(\ref{fitting-function}) 
we obtained an analytic expression for $\chi_{\rm loc}(i \omega_n)$ 
that we used to analyze the numerical results.

In the parameter region 
 $T < 0.1\;T^0_K$, $1 < I/T^0_K < 1.25$ all our data can be fitted with
a {\it single}
value of $\alpha = 0.72 \pm 0.01$ and $\Lambda = 1.54\;T^0_K$. 
The error bar represents the amplitude of variation of $\alpha$
when it is allowed to {\it freely} adjust for each value of $I$
and $T$ in the critical region.
The fits
 are of excellent quality as shown in Fig.~\ref{fig2} where we display 
 results obtained for 
$I = 1.1875\;T^0_K$ at $T=0.011\;T^0_K$. The fitted values of
 $\chi_{\rm loc}(\omega_n=0)$ are represented by the solid lines 
in Fig.~\ref{fig1}.
We found in addition that, within this range of values of $I$ and $T$,  
the fitted $\delta(I,T)$ can be described by the phenomenological
expression  
$\delta \propto [\delta_0(I) + \sqrt{\delta^2_0(I) + 4(T/T^0_K)^2}]/2$
that can be derived using Eq.~(\ref{fitting-function}) in the 
normalization condition (\ref{constraint}).
The parameter $\delta_0$, that decreases linearly with increasing $I$, 
allows us to 
determine the location of the QCP from
the criterion $\delta(I_c,T=0) = 0$: $I_c \approx 1.2\;T^0_K$.

\begin{figure}[t]
\centerline{\epsfxsize=80mm\epsfbox{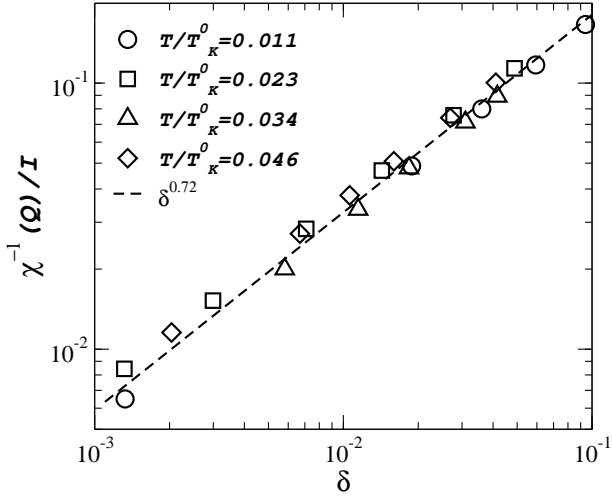}}
\vspace{1ex}
\caption{%
Scaling plot of $\chi^{-1}(Q)$ near
the QCP.}
\label{fig3}
\end{figure}

We have also computed  $\chi^{-1}({\bf Q})$,
the inverse of the static peak susceptibility.
Fig.~\ref{fig3} shows that
$\chi^{-1}({\bf Q}) \propto \delta^{0.72}$ 
for $I$ and $T$ in the quantum critical regime.
Our numerical value for $\alpha$~\cite{universal} is very close
to that seen~\cite{Schroder2} in the Ising-like system 
${\rm CeCu_{6-x}Au_x}$.
The region $I>I_c$ will be discussed elsewhere\cite{ordered}.

We stress
that a simultaneous treatment of 
the Kondo and RKKY
couplings 
is crucial for
our conclusion that the LCP solution is indeed self-consistent. 

Some questions have recently been 
raised concerning the matching of 
the logarithmic terms in the LCP solution~\cite{lcp-long,Burdin} in 
the isotropic model.
Burdin {\it et al.}~\cite{Burdin}
carried out large-$N$ and numerical analyses for 
zero Kondo coupling. They found that the spin liquid (SL) 
phase~\cite{Sachdev-Ye,Parcollet-Georges}  is unstable at low $T$ 
and conjectured that the LCP solutions may not be self-consistent either.
Since the SL phase  corresponds to the {\it stable fixed point} 
of the effective impurity
model~\cite{Zhu} this analysis covers a different parameter
regime and is complementary to ours. By directly accessing the unstable fixed 
point, our 
results 
establish that the logarithmic
terms are self-consistent in the Ising case.
Whether numerical and analytical (beyond RG) studies in the isotropic
case will yield a self-consistent LCP similar to what we have shown
here for the anisotropic model is left for future work.


In summary, we have
numerically identified 
a LCP solution in an anisotropic Kondo lattice model.
The exponent for the ${\bf q}-$dependent dynamical
susceptibility 
is 
fractional
and is close to the experimental
value.

We would like to thank 
S. Burdin, M. Grilli, M. J. Rozenberg (D.R.G.),
K. Ingersent, S. Rabello, J. L. Smith, L. Zhu (Q.S.)
and J. Zhu for 
discussions and collaborations on related problems,
A. Georges, G. Kotliar,
A. Rosch, S. Sachdev and M. Vojta for useful discussions,
and NSF (Grant No.\ DMR-0090071),
Research Corporation,
Robert A.
Welch Foundation, and 
TCSAM
(Q.S.)
for support.
Q.S.\ also acknowledges the hospitality
of
SPhT-CEA/Saclay
and Aspen Center
for Physics.


\begin{references}
\bibitem{universal}
We have also studied three other parameter sets for which 
$  N_0 T^0_K$ = 0.03, 0.085, and 0.25.
The exponent $\alpha$ turns out to be 
essentially the same
suggesting that,
for the Ising-like model,
$\alpha$ is nearly universal.
\end{references}
\end{document}